\documentclass[twocolumn]{aastex63}

\usepackage{graphicx}
\newcommand{\msolar} {$\rm{M_{\odot}}~$}
\newcommand{\msolarc} {$\rm{M_{\odot}}$}
\newcommand{\molH} {$\rm{H_2}$~}
\newcommand{\molHc} {$\rm{H_2}$}

\newcommand{\change}[2][]{%
\ifthenelse{\isempty{#2}}{{\color{red}{#1}}}%
{{\color{orange}\sout{#1}}{\color{red}{#2}}}%
}
\shorttitle{Some First Stars Were Red}
\shortauthors{Woods et al.}
\graphicspath{{./}{figures/}}

\begin{document}

\title{Some First Stars Were Red: Detecting Signatures of Massive Population III Formation Through Long-Term Stochastic Color Variations}

\correspondingauthor{Tyrone E. Woods}
\email{tyrone.woods@nrc-cnrc.gc.ca. tewoods.astro@gmail.com}

\author[0000-0003-1428-5775]{Tyrone E. Woods}
\affiliation{National Research Council of Canada, Herzberg Astronomy \& Astrophysics Research Centre, 5071 West Saanich Road, Victoria, BC V9E 2E7, Canada}
\affiliation{Monash Centre for Astrophysics, School of Physics and Astronomy, Monash University, VIC 3800, Australia}
\author[0000-0002-4201-7367]{Chris J. Willott}
\affiliation{National Research Council of Canada, Herzberg Astronomy \& Astrophysics Research Centre, 5071 West Saanich Road, Victoria, BC V9E 2E7, Canada}
\author[0000-0001-9072-6427]{John A. Regan}
\affiliation{Department of Theoretical Physics, Maynooth University, Maynooth, Ireland}
\author[0000-0003-1173-8847]{John H. Wise}
\affiliation{Center for Relativistic Astrophysics, Georgia Institute of Technology, 837 State Street, Atlanta, GA 30332, USA}
\author[0000-0002-7639-5446]{Turlough P. Downes}
\affiliation{Centre for Astrophysics \& Relativity, School of Mathematical Sciences, Dublin City University, Glasnevin, D09 W6Y4, Ireland}
\author[0000-0002-6622-8513]{Michael L. Norman}
\affiliation{Center for Astrophysics and Space Sciences, University of California, San Diego, 9500 Gilman Dr, La Jolla, CA 92093}
\author[0000-0002-2786-0348]{Brian W. O'Shea}
\affiliation{Department of Computational Mathematics, Science, and Engineering, Michigan State University, MI, 48823, USA}
\affiliation{Department of Physics and Astronomy, Michigan State University,MI, 48823, USA}
\affiliation{Joint Institute for Nuclear Astrophysics - Center for the Evolution of the Elements, USA}
\affiliation{National Superconducting Cyclotron Laboratory, Michigan State, University, MI, 48823, USA}

\begin{abstract}
\noindent Identifying stars formed in pristine environments (Pop III) within the first billion years is vital to uncovering the earliest growth and chemical evolution of galaxies. Pop III galaxies, however, are typically expected to be too faint and too few in number to be detectable by forthcoming instruments without extremely long integration times and/or extreme lensing. In an environment, however, where star formation is suppressed until a halo crosses the atomic cooling limit (e.g., by a modest Lyman-Werner flux, high baryonic streaming velocities, and/or dynamical heating effects),
primordial halos can form substantially more numerous and more massive stars. Some of these stars will in-turn be accreting more rapidly than they can thermally relax at any given time. 
Using high resolution cosmological zoom-in simulations of massive star formation in high-z 
halos, we find that such rapidly accreting stars produce prominent spectral features which would be detectable by {\it JWST}. 
The rapid accretion episodes within the halo lead to stochastic reprocessing of 0--20\% of the total stellar emission into the rest-frame optical over long timescales, a unique signature which may allow deep observations to identify such objects out to $z \sim 10-13$ using mid- and wide-band NIRCam colors alone.

\end{abstract}





\section{Introduction} \label{sec:intro}

\noindent The first stars to have formed in the Universe, though not yet detected, are canonically expected to have been extremely blue, with effective temperatures T~$\sim 10^{5}$K \citep{s02, Murphy_2021}. This is due to their high masses and thus luminosities \citep{mln10}, together with their compactness, which arises from the absence of metal-line opacity in their envelopes \citep{tumlinsonshull00}. These high temperatures, together with their high formation redshifts (z $\sim$ 10 -- 20) and the low efficiency of star formation expected in typical primordial halos, have conspired to temper expectations such ``Population III'' (Pop III) stars may be detected in pristine environments without prohibitively long exposures \citep[e.g.,][]{Zackrisson11}.

However, while the first Pop III galaxies may go undetected, initial suppression of star formation in later primordial halos can ultimately lead to the formation of unusually massive
Pop III stars \citep[e.g.,][]{bl03}. The first Pop III stars to form did so in the shallow potential wells of the first virialised dark matter halos, with masses of $\sim10^5$--$10^6$ \msolarc \citep{abn02}.
In these
early structures, gas cools and contracts under the influence of cooling via vibrational and rotational transitions of \molHc\ \citep{Tegmark_1997}. However, \molH can be readily dissociated by Lyman-Werner (LW) radiation, suppressing initial star formation until the host halo reaches the atomic cooling limit \citep{Field_1966,har00}. Cooling and consequently gravitational collapse can then be triggered by neutral hydrogen once the gas temperature reaches approximately 8000 K. A number of other mechanisms have also been found to suppress initial star formation and permit the formation of very massive or even supermassive stars, with individual stellar masses up to $10^5$ \msolarc\ \citep{tyr17, tyr20a, lupi21}, including strong offsets between dark matter and baryonic streaming velocities within an environment \citep{th10,Tanaka2014, srg17}, or strong dynamical heating either via major mergers \citep{may10} or the rapid assembly of dark matter halos 
\citep[e.g.,][]{wise19}.

Recent numerical simulations have shown that a modest LW background, together with dynamical heating from minor and major mergers, can produce a primordial stellar population which is both much more numerous and more massive than that found in Pop III minihalos \citep{wise19,Regan20}. At the same time, the dense gas flows found in such an environment will lead to intermittent episodes of very rapid ($\gtrsim 0.005 M_{\odot}$/yr) accretion for some stars. This is much faster than the rate at which such stars can thermally equilibrate, and will drive rapid photospheric expansion \citep{kmh77, hle18b}, pushing the effective temperatures of these stars down to $\sim 5\times 10^{3}$ K \citep{hos12} for the duration of the rapid accretion phase. Therefore, in addition to a greatly enhanced total mass formed in stars, we may also expect stochastic, and in some instances very strong, enhancements in the rest-frame optical continuum of these irradiated primordial halos, in particular at wavelengths longer than $\sim $4000 \AA. This presents a uniquely promising opportunity for understanding primordial-composition stars in the early Universe, allowing one to not only plausibly search for these stellar populations themselves but to characterize their formation environments at Cosmic Dawn.

\vskip0.5cm
\section{Numerical Simulations}
\subsection{Halo Models}

In order to compare the stellar populations of primordial halos, we draw on 3D hydrodynamic simulations of two primordial halos described in \citet{Regan20}, which themselves were drawn from the original Renaissance simulations of \cite{Xu13, Xu14} and \cite{ren15}, to which the interested reader is referred to for further details. \textbf{Halo A} crosses the atomic cooling limit before experiencing central gravitational collapse. This halo avoids star formation in its assembly history due to a combination of a mild LW ($\rm{J}_{\rm{LW}} \sim (2$--$10)\times 10^{-21}\rm{erg/s/cm}^{2}\rm{/Hz/sr}$) background and rapid halo assembly, which dynamically heats the halo, suppressing star formation prior to the onset of atomic cooling. \textbf{Halo B}, meanwhile, is not subjected to any LW background, and so star formation is governed by the presence of \molH which allows the gas to cool and collapse, giving rise to a prototypical Pop III mini-halo.

Here, we combine the results of these 3D radiation-hydrodynamics numerical simulations with post-processing using a detailed spectral synthesis code, to produce model spectra for the two fiducial primordial halos {\bf A} and {\bf B}. Using this pipeline, we then produce synthetic photometric magnitudes and colors, and demonstrate how \textbf{Halo A} can be detected and differentiated both from \textbf{Halo B} and more typical high-z galaxies. Finally, we outline detection prospects with forthcoming next-generation facilities such as {\it JWST}.

\label{numerics}
\begin{figure}[t]
    \centering
    \includegraphics[width=0.5\textwidth]{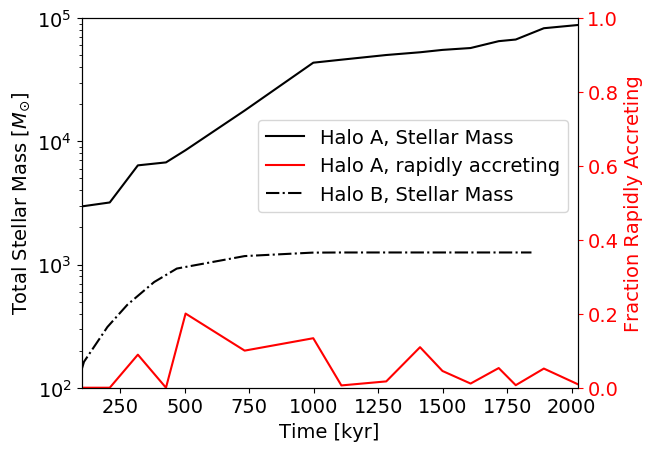}
    \caption{
    Total stellar mass formed (solid, black) and fraction of this mass in rapidly-accreting stars (solid, red) as a function of time for \textbf{Halo A}. For comparison, the total mass formed in \textbf{Halo B} is also plotted (dot-dashed, black), for which no stars are rapidly-accreting.}
    \label{masshistory}
\end{figure}

\subsection{Fragmentation, Rapid Accretion and Massive Star Formation}

For our LW-irradiated model (\textbf{Halo A}), \citet{Regan20} found a $\gtrsim 70\times$ relative increase in the total mass of stars  by the end of the simulation, with a mass spectrum peaking at $\sim 1000\rm{M}_{\odot}$ and reaching $\sim 6000\rm{M}_{\odot}$ (as opposed to stellar masses in the range $\sim$ 10 -- 200 $\rm{M}_{\odot}$ for our control case, \textbf{Halo B}). The final total mass (dark matter, gas and stars) of \textbf{Halo A} was $M_{\rm HaloA} = 9.3 \times 10^7$ \msolar\ (stellar mass: $\approx$ 90,000\msolarc) and the final total mass of \textbf{Halo B} was $M_{\rm HaloB} = 3.7 \times 10^6 $ \msolar\ (stellar mass: $\approx$ 1,300\msolarc).

Critically, the reason {\bf Halo A} stars are unable to reach mass scales of order $\sim10^{5}\rm{M}_{\odot}$, which may however be possible in more strongly-irradiated atomic-cooling halos \citep{lupi21}, is twofold. Firstly, the gas readily fragments due to the presence of \molH \citep{rd18} and secondly the proto-stars only intermittently intersect with pockets of dense infalling gas. The result is that truly rapid accretion ($\gtrsim 5\times10^{-3} \ \rm{M}_{\odot}/\rm{yr}$) cannot be sustained, but arises only at random intervals in the lives of a fraction of the stars formed. The variation of this fraction with time for \textbf{Halo A}, along with the stellar mass evolution of both \textbf{Halo A} and \textbf{Halo B}, are plotted for a series of ``snapshots'' in Fig. \ref{masshistory} (no stars enter such rapid accretion phases in \textbf{Halo B}). As noted in \citet{Regan20}, rapid growth will still greatly inflate the photospheres of those stars which are accreting. This has profound consequences for the anticipated spectra of these primordial stellar populations.

\subsection{Post-processing and Spectral Synthesis}

For each snapshot, we assume that any stars not accreting above the ``inflated photosphere'' threshold \citep[$\sim5\times 10^{-3} \ \rm{M}_{\odot}$/yr,][]{hle18b}, are fully thermally relaxed. In this case, their spectral emission and temperatures are well-described by blackbody emission with an effective temperature of $\approx 10^{5}$K \citep{s02, tyr20a}. 
Stars growing at rates above this threshold, however, accrete much faster than they can thermally relax; the result is that their outer envelopes inflate dramatically \citep{kmh77, hos12}, pushing their effective temperatures down to $\sim$5000K \citep{hos13}.
Here we assume all stars are simply either inflated or not at each snapshot based on their instantaneous accretion rate, given that the intervals between each snapshot are always much longer than the thermal timescale of any accreting star.

All of the stars we consider are massive (up to 1000's of $\rm{M}_{\odot}$); for simplicity, in the following we therefore assume that all stars in either halo radiate at approximately the Eddington limit, such that their luminosity is given by 

\begin{equation}
    L \approx 1.8\times 10^{38}\times (\rm{M}_{*}/\rm{M}_{\odot})~erg/s
    \label{Eddington}
\end{equation}

\noindent Note that in practice, this may overestimate the luminosities for stars in \textbf{Halo A} by up to a factor of $\sim2$, and by significantly more for the typically lower mass stars in \textbf{Halo B} \citep{s02}. As shown below, however, \textbf{Halo B} remains too faint to be detected regardless, therefore this does not significantly impact our results.

Once formed, these stars remain in a dense, gas-rich environment, and the effects of radiative transfer must be accounted for in order to produce a realistic model of their integrated emission. Here, we use the detailed plasma simulation and spectral synthesis code {\sc cloudy}
\citep[version 17.02,][]{cloudy17}, which solves for thermal and ionization equilibrium for arbitrary gas distributions, abundances, and source spectra in 1D. Although the restriction to 1D prevents us from precisely accounting for the clumpy medium seen in both halos, our focus on H and He lines within a modest density range, in an environment which is ionization-bounded, prevents this from significantly impacting our results. 
We assume primordial composition \citep{Planck_2014} and spherical symmetry, and derive ambient density profiles from our source simulations by taking spherical averages of the gas density at 0.5 pc intervals surrounding each star particle \citep[see][for further details]{Regan20}. The spectrum of each star as filtered through its surrounding medium is then computed assuming dynamical and ionization equilibrium, and the resulting emission (including emission from the surrounding medium) is summed for all stars within each snapshot for \textbf{Halo A} and \textbf{Halo B}, respectively.

\section{Observational Prospects}

\subsection{Synthetic Spectra}

\begin{figure}
    \centering
    \includegraphics[width=0.5\textwidth]{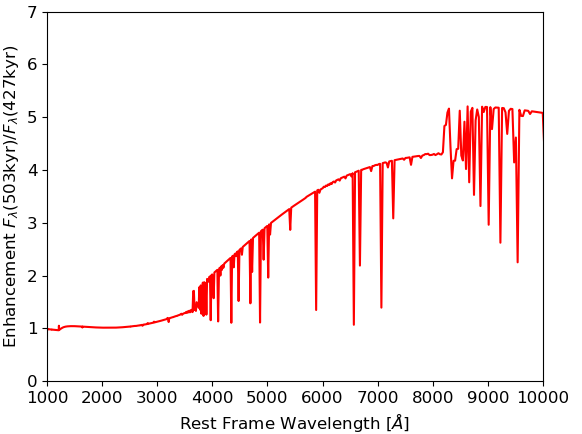}
    \caption{Enhancement in the rest-frame UV, optical, and near-IR spectral energy distribution of our model halo (``Halo A'') at 503 kyr (when many stars are rapidly-accreting) relative to the case at 427 kyr (a quiescent moment). 
    }
    \label{redboost}
\end{figure}

\noindent The total luminosity of \textbf{Halo A} grows throughout the $\sim$2~Myr simulation (up to $\sim$~$10^{43}$~erg/s), tracing the evolution of the total stellar mass (recall eq. \ref{Eddington}) via both new star formation and accretion onto earlier-formed stars. The luminosity of \textbf{Halo B}, however, plateaus at $\sim$~$10^{41}$~erg/s after approximately 500~kyr (at which point both star formation and significant accretion end in \textbf{Halo B}). Both \textbf{Halo A} and \textbf{Halo B} exhibit comparably prominent hydrogen and helium emission lines, as expected for Pop III galaxies \citep{s02}. For instance, both halos exhibit rest-frame optical He II 4686\AA/H$\beta$ ratios that typically fall around $\sim$0.06, albeit with some very modest departures consistent with variations in the gas temperature and density seen between the two halos. This is consistent with the comparable ratios of H (13.6 eV) and He$^{+}$ (54.4 eV) ionizing photons between the two halos, although note that the ratio of ionizing photons to the mass of new stars formed, and thus the total luminosity of either line, varies stochastically in {\textbf{Halo A}} in response to the occurrence of rapid-accretion episodes for some stars (see below).

Other than the rapid growth in \textbf{Halo A's} luminosity, the most striking difference between \textbf{Halo A} and \textbf{Halo B} is the stochastic enhancement in the optical continuum of \textbf{Halo A} driven by the sporadic rapid accretion episodes encountered by some of its stars. The net effect of accounting for the resulting photospheric expansion is that a variable but often significant fraction of the total luminosity of the halo (between 0 -- 20\% in the case of our \textbf{Halo A} simulation) is reprocessed from the extreme UV to the optical. This may be contrasted with e.g., the typically $\sim$3\% of any $T \sim 10^{5}$K object's luminosity which is reprocessed into H$\alpha$, the strongest rest-frame optical line, under the conditions prevailing in our simulations. The emergent spectrum from these rapidly-accreting objects 
contributes to a broad continuum enhancement, particularly at wavelengths above $\sim$~4000\AA. A relatively extreme example of this is shown in Fig.~\ref{redboost}, where a $\sim$5-fold continuum enhancement is seen early in the evolution of \textbf{Halo A} as it transitions from a quiescent moment to a resumption of rapid growth. 


\begin{figure}
    \centering
    \includegraphics[width=0.5\textwidth]{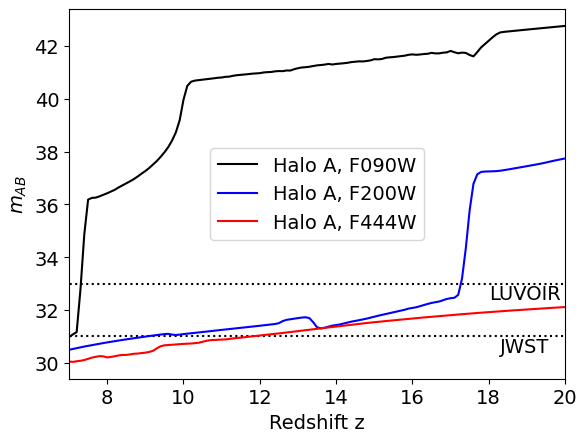}
    \caption{Wide band AB magnitudes for the F200W (blue) and F444W (red) NIRCam filters shown as a function of assumed redshift, for Halo A at 1892 kyr (note that here and throughout, we take a single simulation and redshift its rest frame spectrum accordingly). Also shown for reference is the Lyman-limit drop-out above z$\sim$7, as seen in the F090W band (black). Horizontal dotted black lines denote a marginal detection with a S/N$\sim$5 for {\it JWST} and {\it LUVOIR} after $\sim17$ and $\sim$10 hour integrations, respectively.}
    \label{magnitudes}
\end{figure}


\subsection{Photometric Magnitudes and Colors with Redshift}

\noindent In the near-term, the ideal instrument for accessing the rest-frame optical at high redshifts, with sufficient sensitivity and spatial resolution, is the {\it JWST}.
In order to produce synthetic observed magnitudes based on our model spectra, we use the software package {\sc synphot}\footnote{https://synphot.readthedocs.io/en/latest/} \citep{synphot}. Note that while the simulations of both \textbf{Halo A} and \textbf{Halo B} found the onset of star formation to arise at $z\sim$15, a broad range of formation redshifts are possible depending on the availability of metal-free gas in an appropriate environment \citep{regan19}.

The simulated F200W and F444W magnitudes near the end of the simulation for \textbf{Halo A} 
are shown in Fig.~\ref{magnitudes}, 
as a function of the assumed formation redshift (as well as the F090W band, which illustrates the Lyman-limit drop-out for $z \gtrsim 7$). Remarkably, the strongly enhanced star formation efficiency and stellar accretion within \textbf{Halo A} renders it detectable at modest redshifts, given long yet plausible integration times, e.g., for a F444W AB magnitude = 31, NIRCam would reach a S/N$\sim$5 in a 17 hour observation in a low background field ({\it JWST} Exposure Time Calculator\footnote{https://jwst.etc.stsci.edu}), allowing {\it JWST} to detect objects like \textbf{Halo A}  out to $z\sim12$. Future proposed missions, such as the {\it Large UV/Optical/Infrared Surveyor (LUVOIR)} mission concept, would be able to extend this to $z\sim17$, reaching a planned S/N$\sim$5 in $\sim$10 hours \citep{LUVOIR}. This is in stark contrast with \textbf{Halo B}; even when assuming the Eddington luminosity for all stars,
\textbf{Halo B} only reaches $m_{\rm{AB}}\sim 35$ by the end of its evolution assuming formation at $z\sim7$. 
Typical for Pop III halos, then, \textbf{Halo B} would only be detectable, even given next-generation facilities, with the benefit of extreme lensing \citep[e.g.,][]{zet12}.

\begin{figure}
    \centering
    \includegraphics[width=0.5\textwidth]{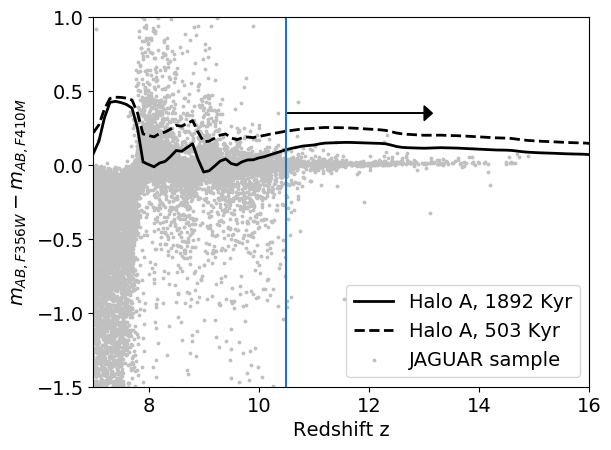}
    \caption{Observed NIRCam F356W - F410M color of \textbf{Halo A} as a function of the redshift assumed for snapshots at 503 kyr (dashed; rapidly accreting) and 1892 kyr (solid; final). 
    Shown for comparison is the same NIRCam color for 
    the JAGUAR mock catalog of expected JWST high-z sources. The vertical bar highlights the approximate redshift (z$\sim$10.5) above which stochastic reddening can provide a distinct color signature of rapidly-accreting Pop III stars.}
    \label{color_redshift}
\end{figure}

The faintness and uncertain number density of objects like \textbf{Halo A}, however, necessitates a search strategy that does not initially rely on time-intensive high-S/N spectroscopy. The question then arises whether such objects 
could be identified by their broad- and medium-band photometric colors alone. 
The dense inflowing gas and stochastically-arising rapid accretion episodes within \textbf{Halo A} lead to pronounced episodes of reddening over the lifetime of the starburst, in sharp contrast to \textbf{Halo B}.
This suggests that rapidly-accreting Pop III stars may be distinguished from more typical Pop III minihalos based on color alone. 
Shown in Fig.~\ref{color_redshift} is the color of two snapshots of \textbf{Halo A} 
plotted as a function of redshift.
For comparison, we also plot the same colors for simulated star forming galaxies from the JAdes extraGalactic Ultradeep Artificial Realizations (JAGUAR) package \citep{JAGUAR}.

The redder continuum of these models shows up as a broad excess  approximately 0.1--0.3~magnitudes redder than the locus of the JAGUAR models in a number of {\it JWST} NIRCam colors, e.g. at z~$\sim$~10.5--13 for F356W - F410M. Note that differentiating colors to such a precision would require a $\sim 10\sigma$ detection, necessitating either extremely long integrations or more modestly deep observations of lensed environments. Critically, since the majority of \textbf{Halo A}'s emission remains in the blue, such objects could still be identified as high-z (and in particular, disambiguated from brown dwarfs and other low-z objects) based on their Lyman drop-out (see, e.g., the predicted F090W magnitudes in Fig. \ref{magnitudes}, which show a Lyman-limit cutoff for z$\gtrsim 7$).
At lower redshifts ($7 \lesssim z \lesssim 8$), the narrower, deeper feature opposite to the JAGUAR locus is due to H-beta being within the filter for the model halo, whereas there is stronger [OIII] 5007\AA\ emission in the neighbouring filter for the JAGUAR mock galaxies; such behaviour is generic for any metal-free stellar population with strong Balmer lines \citep{Zackrisson11}.

\section{Discussion}


Although it is clear objects such as \textbf{Halo A} may be detectable by {\it JWST}, the question remains whether or not they will be sufficiently numerous to be found given {\it JWST}'s relatively small footprint. Here, we may make some indicative estimates from comparison with previous studies of supermassive star formation rates \citep[see e.g.,][for a review]{titans}. The critical conditions needed for supermassive star formation in primordial halos were long approximated as a single threshold Lyman-Werner intensity within the ambient environment 
\citep{bl03} although recent work suggests a realistic treatment is considerably more nuanced \citep[e.g.,][]{Sugimura14,Agarwal16,Wolcott2017}.
Estimates for the critical Lyman-Werner intensity needed for supermassive star formation have varied widely, with consequently wide variation in their expected number densities e.g., for $\rm{J}_{\rm{crit}}\sim30~\rm{J}_{\rm{LW}}$, \cite{agarw12} find $\sim 10^{-2}$--$10^{-1}~\rm{cMpc}^{-3}$ at $z \sim 10$, whereas \cite{dijkstra08} found only $\sim10^{-6}~\rm{cMpc}^{-3}$ for $\rm{J}_{\rm{crit}}>1000~\rm{J}_{\rm{LW}}$. A number of other factors, however, including dynamical heating \citep[e.g.,][]{wise19} and baryonic streaming velocities \citep[e.g.,][]{Tanaka2014}, may also contribute to suppressing typical star formation in primordial halos, providing complementary means for seeding supermassive stars which may imply greater number densities.

The conditions needed for simply enhancing the total mass, individual stellar masses, and stochastic accretion rates within Pop III halos, as seen above, appear to be similar if more modest. In the case of \textbf{Halo A}, the key factor was found to be dynamical heating caused by the rapid assembly of the halo, more than the moderate ambient LW background \citep{wise19, Regan20}. This rapid assembly heats the halo, similarly delaying and suppressing star formation. \cite{wise19} calculated the number density of such halos to be approximately $\sim 1 \times 10^{-3} \ \rm{cMpc}^{-3}$ at $z \sim 15$, suggesting an occurence rate as high as $\sim$ a few per NIRCam field of view (for $\Delta z$ = 1). The region in which the parent simulations were run, however, was an overdense region \citep[see for example,][]{Xu_2013}, and when this is accounted for the total expected number density over the whole sky falls to $\sim 1 \times 10^{-6} \ \rm{cMpc}^{-3}$ at $z \sim 15$ \citep[e.g., $\sim$ 1 per 200 NIRCam pointings, and consistent with][]{dijkstra08}. The formation rate of such objects in ``normal'' (as opposed to overdense) regions is expected to rise significantly at lower redshifts, however, at least up to $z \sim 10-12$ \citep[e.g.,][]{regan19}; beyond this point, increasing metal pollution may hinder the rise of pristine atomic-cooling halos at still lower redshifts, though this remains poorly constrained.

Ultimately, our analysis represents only a single simulation, and the real abundance of similarly-enhanced Pop III halos cannot be reliably estimated without a greater understanding of the relative importance of LW- irradiation \citep[including in synchronised halos, e.g.,][]{regan17a}, dynamical heating from mergers, and baryonic streaming velocities in producing such populations in average-density regions of the early Universe, which is beyond the scope of the present work. We may argue, however, that depending on the real evolving number density as a function of redshift:

\begin{itemize}
    \item If objects such as Halo A continue to form up until z$\sim$7--8, then planned observations such as the JADES GTO program \citep[whose deep survey will reach a S/N$\sim$10 for AB mag $\sim$30,][]{JAGUAR} will be able to detect similar objects and identify them by their color.
    
    \item If Halo A-like objects only form at z$\gtrsim$10, then robustly detecting them without the benefit of lensing would be too time-intensive for blind surveys (e.g., reaching S/N$\sim$10 for z$\sim$12 objects would require $\sim$68 hours on JWST). Shallow or pure parallel surveys with JWST, however, should soon reveal overdensities at z$\sim$10, as will deep surveys with e.g., the Roman Space Telescope \citep{Vikaeus21}. These overdensities may then be targeted for deep follow-up observations.
    
    \item If the true number density of Halo A-like objects is significantly greater than our estimate based on expected supermassive star formation sites, it may be viable to search for lensed objects even in untargeted surveys. For example, we note that in the  CANUCS survey, we expect ~10 sq. arc min. within its planned footprint to have a magnification $\gtrsim$3--4 (Willott, private communication). This suggests such an approach may be able to detect (or rule out) formation rates of Halo A-like objects consistent with the very highest number densities of such halos.
    
    \item Finally, if the formation rate of Halo A-like objects is significantly lower than our estimates given above, or if they arise only at higher redshifts (e.g., z$\gtrsim$15), such objects may yet be revealed by next generation facilities such as the proposed LUVOIR mission (recall Fig. \ref{magnitudes}). 
\end{itemize}

\section{Conclusion}
\noindent Here we have shown that rapidly-accreting Pop III stars, arising in pristine environments wherein star formation was initially suppressed, may have a profound influence on the spectral features and color of some primordial stellar populations. In particular, chance encounters between Pop III stars and dense gas within such halos will lead to rapid accretion episodes which will greatly inflate the photospheres of these stars, potentially re-processing a significant fraction of the halo's total stellar emission to the rest-frame optical. Together with the highly-enhanced star formation indicated by our models in such environments, we have shown that such populations would be detectable by forthcoming next-generation infrared facilities, and distinguishable from more typical, metal-enriched stellar populations at $z \gtrsim 10.5$ and at $7 \lesssim z \lesssim 8$ based on their broad- and medium-band colors alone. Conversely, this stochastically-arising reddening of some Pop III halos may provide an invaluable probe of the star formation history and environment in overdense regions of the early Universe, providing a direct probe of the total mass in rapidly-accreting and quiescent states within such early populations.


\acknowledgments

The authors would like to thank the anonymous referee for their careful reading and helpful comments. TEW acknowledges support from the National Research Council Canada's Plaskett Fellowship. JR acknowledges support from the Royal Society and Science Foundation Ireland under grant number URF$\backslash$R1$\backslash$191132. JHW is supported by National Science Foundation grants AST-1614333 and OAC-1835213, and NASA grants NNX17AG23G and 80NSSC20K0520.  BWO acknowledges support from NSF grants PHY-1430152, AST-1514700, AST-1517908, and  OAC-1835213, by NASA grants NNX12AC98G and NNX15AP39G, and by HST-AR-13261 and HST-AR-14315. JR wishes to acknowledge the DJEI/DES/SFI/HEA Irish Centre for High-End Computing (ICHEC) for the provision of computational facilities and support on which the zoom simulations were carried out.


\bibliography{sample63}{}
\bibliographystyle{aasjournal}

\end{document}